\documentclass[12pt]{article}
\usepackage[utf8]{inputenc}
\usepackage{fancyhdr}
\usepackage{amsmath}
\usepackage{amsfonts}
\usepackage{mathrsfs}
\usepackage{graphicx}
\usepackage{amssymb}
\usepackage{hyperref}

\begin{document}

\title{$\beta$-function for topologically massive gluons}
\author{{Debmalya Mukhopadhyay}\thanks{debmalya@bose.res.in}\,
 and
  \ {Amitabha Lahiri}\thanks{amitabha@bose.res.in} \ \\\\ 
\textit{{\small S.N. Bose National Centre of Basic Science,}}\
\textit{{\small  Sector-III, Block-JD,}}\\
\textit{{\small Saltlake,  Kolkata, West Bengal}}\\
\textit{{\small India.}}}
\date{\today}
\maketitle

\begin{abstract}
We calculate the quantum corrections to the two-point function
of four dimensional topologically massive non-Abelian vector 
fields at one loop order for $SU(N)$ gauge theory in 
Feynman-'t~Hooft gauge. We calculate the beta function of the 
gauge coupling constant and find that the theory becomes 
asymptotically free faster than pure $SU(N)$ gauge theory. 
  \end{abstract}

\section{Introduction}
The recent discovery of a 125-GeV Higgs boson~\cite{Aad:2012tfa,
  Chatrchyan:2012ufa} has completed the observation of the
fundamental particles of the Standard Model~\cite{Weinberg:1967tq,
  Salam:1968rm}.  But it has not completed the description of the
low energy particle universe.  The mechanism for neutrino masses
remains incompletely understood, as does the mechanism of family
symmetry breaking. There may be more particles lurking in the
shadows not yet illuminated by the LHC, if not superpartners, then
at least the particles that make up dark matter.

The mechanism of color confinement also remains unknown. It is not
clear if the LHC will be able to shed any light on this problem,
since confinement occurs at low energy, and quarks and gluons are
effectively free particles at the energies probed by the
LHC. However, the LHC may be able to answer a related question,
that of whether strong interactions, which are short-range like the
weak interactions, are also mediated by massive vector bosons. The
idea of a dynamically generated gluon mass has been at the root of
much recent activity (see e.g.~\cite{Aguilar:2013hoa} and
references therein). It has been known for a long time that such a
mass provides a qualitative understanding of many dimensionful
parameters of QCD, including the string
tension~\cite{Cornwall:1981zr}. In fact a gauge-invariant mass of
the gluon is just as useful~\cite{Cornwall:1979hz}, and plays an
important role in the center-vortex picture of
confinement~\cite{Cornwall:1997ds}. In this picture, vortices of
thickness $\sim m^{-1}$\,, and carrying magnetic flux in the center
of the gauge group, are assumed to form a condensate, where $m$ is
the mass of the gauge boson. For fundamental Wilson loops that are
large compared to $m^{-1}$ an area law can be shown to arise,
although loops of size $\sim m^{-1}$ obey a perimeter law. However,
unbroken Yang-Mills theory with massive gluons is generally
believed to be not renormalizable.

In this paper we consider a special kind of gauge-invariant mass,
namely topologically generated gluon mass, which generalizes the
similarly named Abelian mechanism~\cite{Allen:1990gb}.  In this
model, the field strength $F$ of the gauge field is coupled to an
antisymmetric tensor potential $B$ via a term of the form
$\epsilon^{\mu\nu\rho\lambda}\mathrm{Tr}\,
B_{\mu\nu}F_{\rho\lambda}$\,. A kinetic term for the $B$ field is
also included, leading to the gauge field propagator developing a
pole. The model does not require spontaneous symmetry breaking, can
be shown to be unitary~\cite{Hwang:1995er, Lahiri:1996dm,
  Lahiri:2011ic}, and there is good reason to believe that it is
also renormalizable in an algebraic sense~\cite{Lahiri:1999uc}, but
it is not known if this massive theory remains asymptotically free.
In this paper we investigate how the gauge field propagator is
modified at one loop by the coupling with the tensor field, and
calculate the beta function of gauge coupling constant.

In pure Yang-Mills theory with gauge group $SU(N)$, the coupling 
constant at one loop runs with energy as
\begin{eqnarray}
\alpha(Q^2)=\frac{\alpha(\mu^2)}{1+N\frac{11}{12\pi}
\alpha(\mu^2)\ln\frac{Q^2}{\mu^2}}\,,
\end{eqnarray}  
where $\mu$ is the renormalization point, and $\alpha$ is related
to the gauge coupling constant $g$ by $\alpha=\frac{g^2}{4\pi}\,.$
The corresponding beta function is then
     \begin{eqnarray}
     \beta(\alpha)=-\frac{11}{3}N \frac{\alpha^2}{2\pi}\,.
\end{eqnarray}  
When interactions with fermions or scalars are included,
the beta function gets modified to
\begin{eqnarray}
\beta(\alpha)=\left(-\frac{11}{3} N+ n N_f
\right)\frac{\alpha^2}{2\pi}\,,
\label{intro.beta}
\end{eqnarray} 
where $n$ is $\frac{2}{3}$ for a fermion, $\frac{1}{3}$ for a
complex scalar and $\frac{1}{6}$ for a real scalar field, and $N_f$
is the number of flavors. The positive sign before the second term
in the expression in Eq.~(\ref{intro.beta}) signifies the screening
effect due to virtual pairs of matter particles.  Thus perturbation
theory breaks down for sufficiently large number of species of
particles, and this equation allows us to calculate the energy
where it does so.

We will see below that the antisymmetric tensor has an 
anti-screening effect on the gauge coupling constant. This is an 
effect of the specific coupling considered, and we should think
of the anti-screening as an effect of the gauge-invariant 
topological mass of the gluon. We find the result surprising, 
since massive gluons are short-range. 

\section{Feynman rules for the model}\label{feyn}
We start from the Lagrangian 
\begin{eqnarray}
\nonumber \mathscr{L}&=& -\frac{1}{4}F^{\mu\nu}_a F_{\mu\nu}^a+
\frac{1}{12}H^{\mu\nu\lambda}_a H_{\mu\nu\lambda}^a+
\frac{m}{4}\epsilon ^{\mu\nu\rho\lambda}F_{\mu\nu}^a
B_{\rho\lambda}^a \\ && \qquad + \partial_\mu
\bar{\omega}_a\partial^\mu \omega^a - g f_{bca} A_\mu^b\partial^\mu
\bar{\omega}^a\omega_c-\frac{1}{2\xi}(\partial_\mu A^\mu_a)^2\,.
\label{feyn.lag}
\end{eqnarray}
Here  $F^{\mu\nu}_a$ is the field strength of the gauge bosons,
\begin{eqnarray}
F^{\mu\nu}_a&=& \partial^\mu A^\nu_a-\partial^\nu A^\mu_a -
gf_{bca} A^\mu_b A^\nu_c\,,  
\end{eqnarray}
and $H^a_{\mu\nu\lambda}$ is the field strength of the second rank
anti-symmetric field $B^{\mu\nu}_a$, given by 
\begin{eqnarray}
H^{\mu\nu\lambda}_a&=&\partial^{[\mu}
B^{\rho\lambda]}_a-gf_{bca}A^{[\mu}_b B^{\rho\lambda]}_c\,, 
\end{eqnarray}
where the square brackets imply sum over cyclic permutations,
$\partial^{[\mu}B^{\rho\lambda]}_a=\partial^{\mu}
B^{\rho\lambda}_a+\partial^{\rho}
B^{\lambda\mu}_a+\partial^{\lambda} B^{\mu\rho}_a$.  The Lagrangian
density has been written in terms of the renormalized fields and
coupling constants.


The quadratic in the kinetic term of $B_{\mu\nu}^a$ cannot be
inverted for obtaining the propagator.  We add a term invariant
under $SU(N)$ gauge invariance, 
\begin{eqnarray}
\mathscr{L}'=\frac{1}{2\eta}(D_\mu B^{\mu\nu}_a)^2\,,
\label{feyn.BGF}
\end{eqnarray}
to the Lagrangian density in Eq.~(\ref{feyn.lag}) to get the
propagator of the tensor field. Here $\eta$ is a arbitrary
parameter and $D_{\mu}$ is the gauge covariant derivative. The
propagators of the fields are from the quadratic parts from the
renormalised Lagrangian density and $\mathscr{L}'$ excluding the
$B\wedge F$ term, are
\begin{eqnarray}
i\Delta_{\mu\nu,ab} &=& -\frac{i}{k^2}\left(
  g^{\mu\nu}-(1-\xi)\frac{k^\mu k^\nu}{k^2} \right) \delta_{ab}
\label{feyn.massless-Aprop}\\ 
i\Delta_{\mu\nu,\rho\lambda;ab}
&=&\frac{i}{k^2}\left(g_{\mu[\rho}g_{\lambda]\nu}+(1-\eta)\frac{k_{[\mu}
    k_{[\lambda} g_{\rho]\nu]}}{k^2}\right) \delta_{ab} 
\label{feyn.massless-Bprop}
\end{eqnarray}

\begin{figure}[tbh]
\begin{center}
\includegraphics[scale=.04]{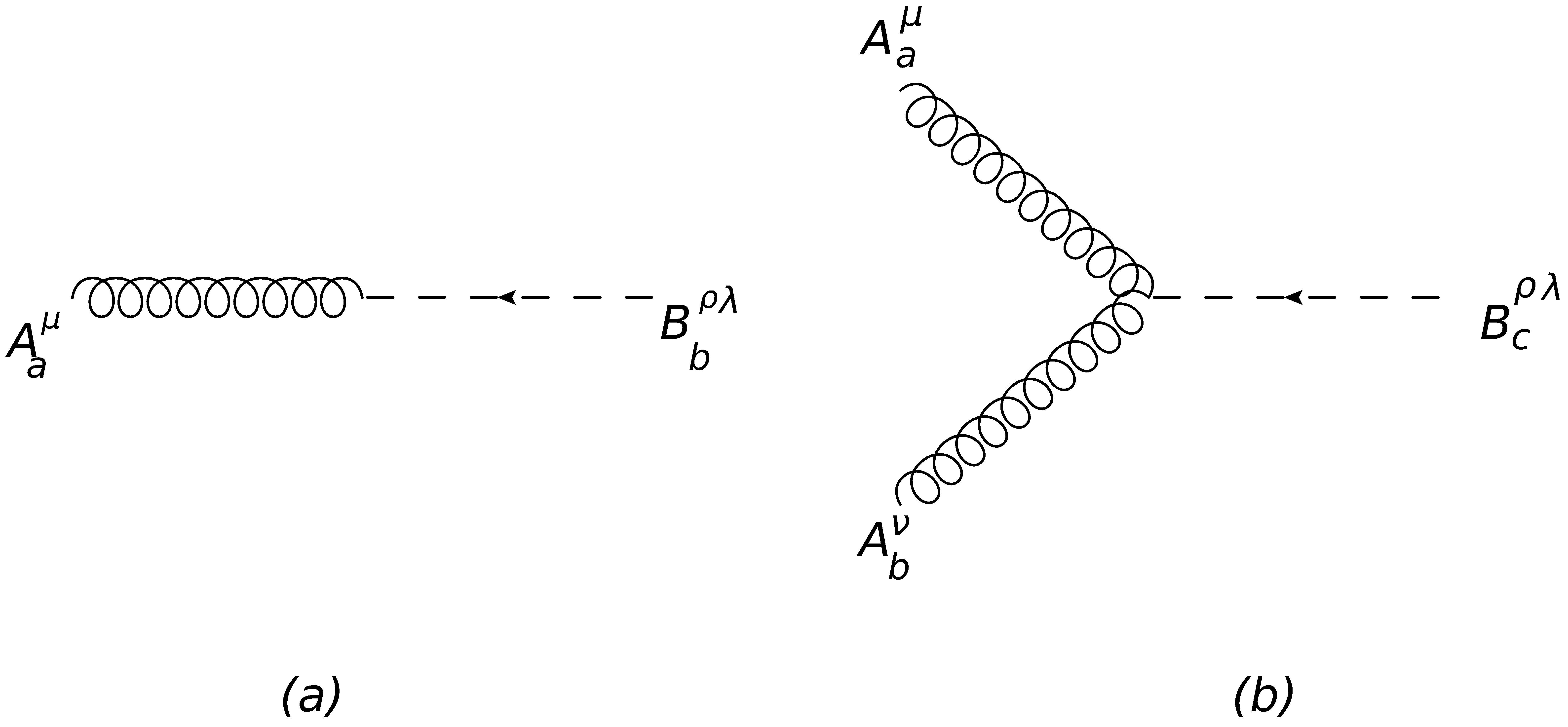}
\end{center}
\caption{(a) Two-point vertex due to the $AB$ interaction; (b)
  three-point vertex due to the $ABB$ interaction.} 
\label{feynfig.AB-AAB}
\end{figure}

We get a two-point coupling of gauge field and $B$ field from the
third term in the Lagrangian of Eq.~(\ref{feyn.lag}). It causes
mixing of the $A$ field with the $B$ field. The third term also
contains an $ABB$ interaction. The vertex rules for the two-point
coupling and the three-point coupling are respectively
\begin{eqnarray}
iV_{\mu\nu, \lambda}^{ab} &=& -m \epsilon_{\mu\nu\lambda\rho}k^\rho 
\delta^{ab}\,, \\ 
iV_{\mu,\nu, \lambda\rho}^{abc} &=& igm
f^{bca}\epsilon_{\mu\nu\lambda\rho}\,. 
\end{eqnarray}
The corresponding vertex diagrams are displayed in
Fig.~\ref{feynfig.AB-AAB}\,. 

\begin{figure}[tbh]
   \centering
\includegraphics[scale=.5]{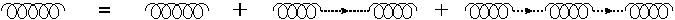}
\caption{Sum over insertions of the $B$-propagator into
  $A$-propagator.} 
\label{feynfig.propagator}
\end{figure}

Because of the two-point coupling, the quadratic terms in the $A$
and $B$ fields are not diagonal, the two-point vertex corresponds
to an off-diagonal mixing term. When calculating the propagator of
the $A$ field, one has to sum over all tree diagrams in which the
$B$ propagator is inserted into the $A$ propagator via this
two-point interaction, as shown in
Fig.~{\ref{feynfig.propagator}}\,.  This `diagonalizes' the matrix
of the propagators.

\smallskip

The complete propagator is given by summing the diagrams in this
infinite series~\cite{Allen:1990gb} 
\begin{eqnarray}
iD_{\mu\nu}^{ab} &=&\left( i\Delta_{\mu\nu} +
i\Delta_{\mu\mu'} iV_{\sigma\rho, \mu'} i
\Delta_{\sigma\rho, \sigma'\rho'} \,
iV_{\sigma'\rho', \nu'} i \Delta_{\nu'\nu} + \cdots \right)\delta^{ab}
\nonumber \\ 
&=&  -i\left[\frac{g_{\mu\nu}-\frac{k_\mu
      k_\nu}{k^2}}{(k^2 - m^2)}+\xi \frac{k_\mu k_\nu}{k^4}\right]
\delta^{ab}\,.
\label{feyn.aprop}
\end{eqnarray} 
Similarly, the full tree-level propagator of the $B_{\mu\nu}$ field
can be obtained summing over insertions of the $A$-propagator via
the two-point coupling, 
\begin{eqnarray}
 iD_{\mu\nu,
  \rho\lambda}^{ab}&=&i\left[\frac{g_{\mu[\rho}g_{\lambda]\nu}+\frac{k_{[\mu}
      k_{[\lambda} g_{\rho]\nu]}}{k^2}}{k^2-m^2}-\eta
  \frac{k_{[\mu} k_{[\lambda} g_{\rho]\nu]}}{k^4}\right]
\delta^{ab}\,.
\label{feyn.bprop}
\end{eqnarray}
We will choose the Feynman-'t~Hooft gauge $\xi=1\,,$ and put
$\eta=1\,,$ to further simplify the calculations.

\begin{figure}[htbp]
 \begin{center}
    \includegraphics[scale=0.03]{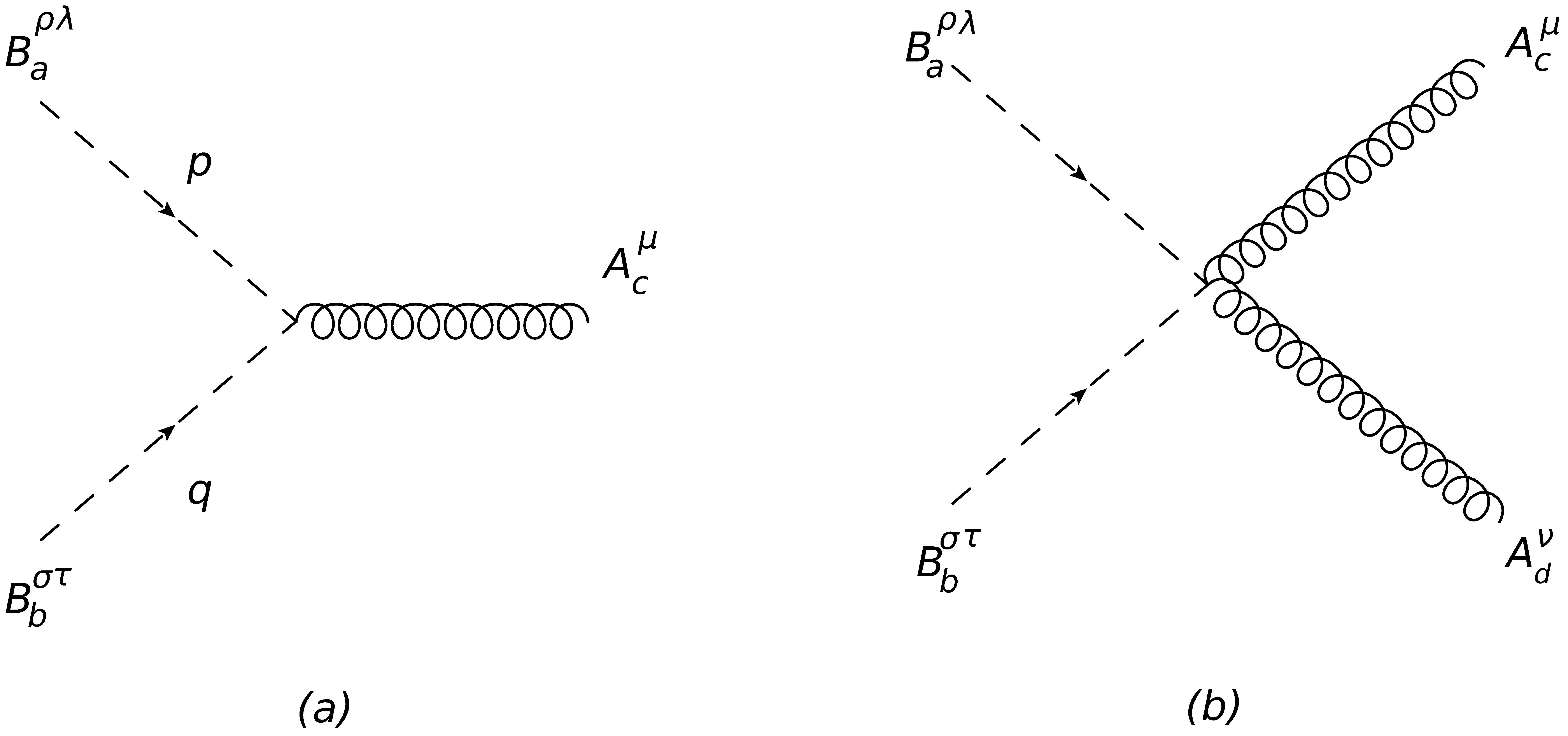} 
\caption{ABB and AABB vertices;}
  \label{feynfig.ABB-AABB}
  \end{center}
\end{figure}

Next we consider the interactions coming from the terms quadratic
in the $B$-field, namely the second term of Eq.~(\ref{feyn.lag}) and
the added term of Eq.~(\ref{feyn.BGF}). These terms contain $ABB$
and $AABB$ interactions, with diagrams shown in
Fig.~\ref{feynfig.ABB-AABB}. 
Their vertex rules are respectively
\begin{eqnarray}
 iV^{abc}_{\mu, \lambda\rho, \sigma\tau} &=&
gf^{abc}\left[(p - q)_\mu g_{\lambda[\sigma}g_{\tau]\rho} +
\left(p+q/{\eta}\right)_{[\sigma}g_{\tau][\lambda}g_{\rho]\mu} 
-\left(q+p/{\eta} \right)_{[\lambda}g_{\rho][\sigma}g_{\tau]\mu}
\right]\,, \nonumber \\
\end{eqnarray}
and
\begin{eqnarray}
\nonumber iV^{abcd}_{\mu, \nu, \lambda\rho, \sigma\tau} &=& ig^2
\left[f_{ace}f_{bde}\left(g_{\mu\nu}g_{\lambda[\sigma}g_{\tau]\rho}+ 
g_{\mu[\sigma}g_{\tau]g[\lambda}g_{\rho]\nu}-\frac{1}{\eta}
g_{\mu[\lambda}g_{\rho][\sigma}g_{\tau]\nu}\right) \right. 
\\  &&+ \left.
f_{ade}f_{bce}\left(g_{\mu\nu}g_{\lambda[\sigma}g_{\tau]\rho} +
  g_{\mu[\lambda}g_{\rho]g[\sigma}g_{\tau]\nu} -
  \frac{1}{\eta}g_{\mu[\sigma}g_{\tau]g[\lambda}
  g_{\rho]\nu}\right)\right]\,.\nonumber \\
\end{eqnarray}

The gauge coupling constant $g$ is a gauge invariant quantity,
hence its variation with the energy, i.e. beta function, does not
depend on the choice of gauge fixing term, in particular the value
of the gauge-fixing parameter $\xi\,.$ The parameter $\eta$ is also
unaffected by gauge transformations, so any value can be chosen for
$\eta$ without afffecting the $\beta$-function. Therefore, as
mentioned above, we choose the Feynman-'t~Hooft gauge $\xi=1\,,$
and also put $\eta=1$ for the simplification of the calculations.

We use dimensional regularization for the loop integrations. The
integrations are done in $4-\epsilon$ dimensions. Since our goal is
to get the beta function at one-loop order, it is sufficient to
calculate the coefficient of $\frac{2}{\epsilon}$ in the result of 
the loop-integration. 

\section{One loop diagrams}\label{old}
\begin{figure}[htbp]
\begin{center}
    \includegraphics[scale=0.06]{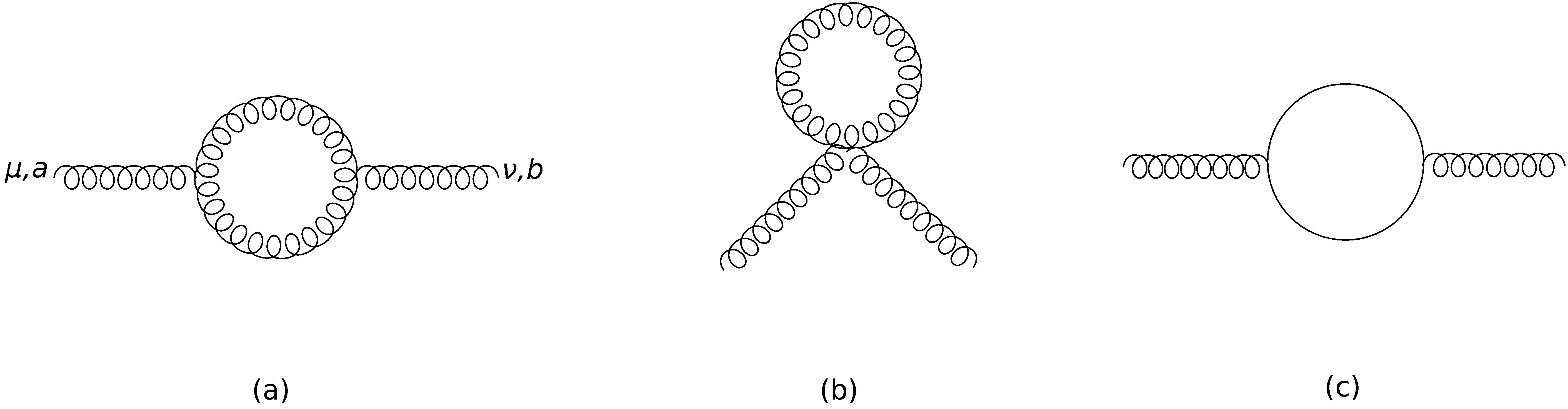} 
  \end{center}
\caption{(a) and (b): Gluon loop; (c): ghost loop;}
  \label{oldfig.glgh}
\end{figure}

In this section we calculate the diagrams which contribute to the
one-loop $\beta$-function for the gauge coupling constant $g\,.$ We
start with the diagrams which appear in pure Yang-Mills
theory. These are generated only by the three and four-point
couplings of the gauge and ghost fields, with the diagrams shown in
Fig.~\ref{oldfig.glgh}. We should check if there are any
differences in the divergent part with the known result in the
literature, since we are now considering massive gauge fields
rather than massless ones. It is easy to see that the divergent
parts of the diagrams in Fig.~\ref{oldfig.glgh}(a) and
Fig.~\ref{oldfig.glgh}(b) are different from the ordinary
(massless) case, whereas the ghost loop in
Fig.~\ref{oldfig.glgh}(c) remains unaffected.
Denoting the four-momentum of the external legs as $p^\mu$, we
calculate that the divergent part for Fig.~\ref{oldfig.glgh}(a) is 
given by 
\begin{eqnarray}
\Pi_{ab,\mu\nu, {\epsilon}}^{\ref{oldfig.glgh}a}=
\frac{1}{2}\,\frac{N\delta^{ab}g^2}{16\pi^2}
\left[\frac{19 g_{\mu\nu}p^2 - 22p_{\mu}p_{\nu}}{6}
-\frac{3}{2}m^2 g_{\mu\nu}\right]\,,
\end{eqnarray}
where the factor of $\frac{1}{2}$ appears due to the presence of
identical gauge boson propagators in the loop. The subscript
$\epsilon$ is placed on $\Pi_{ab,\mu\nu, {\epsilon}}^{n}$ to
indicate that the right hand side of this equation is the
coefficient of $\frac{2}{\epsilon}$. It should be obvious that the
second term containing $m^2$ appears due to the massive pole in the
internal propagator. In fact this is the only difference with the
same diagram for massless gauge bosons.  The divergent part for the
diagram in Fig.~\ref{oldfig.glgh}(b) is calculated to be
\begin{eqnarray}
\Pi_{ab,\mu\nu, {\epsilon}}^{\ref{oldfig.glgh}b}=
-\frac{1}{4}\,\frac{N\delta^{ab}g^2}{16\pi^2}\, 9m^2 g_{\mu\nu}\,.
\end{eqnarray} 
This diagram is known to vanish for massless gauge fields, as is
borne out by the mass dependence of the amplitude calculated here.
The contribution from the divergent part for the diagram in
Fig.~\ref{oldfig.glgh}(c), which contains a ghost loop, does not
change from the usual $SU(N)$ gauge theory because the Fadeev-Popov
ghosts are massless. So we can write
\begin{eqnarray}
\Pi_{ab,\mu\nu, {\epsilon}}^{\ref{oldfig.glgh}c}=
-\frac{N\delta^{ab}g^2}{16\pi^2} 
\frac{(-g_{\mu\nu}p^2-2 p_{\mu}p_{\nu})}{12}\,. 
\label{old.ghprop}
\end{eqnarray}
Adding up the three contributions, we calculate the total
contribution of the loops in Fig.~\ref{oldfig.glgh},
\begin{eqnarray}
\Pi_{ab,\mu\nu, {\epsilon}}^{\ref{oldfig.glgh}} &=&
\Pi_{ab,\mu\nu, {\epsilon}}^{\ref{oldfig.glgh}a}
+\Pi_{ab,\mu\nu, {\epsilon}}^{\ref{oldfig.glgh}b}
+\Pi_{ab,\mu\nu, {\epsilon}}^{\ref{oldfig.glgh}c} \nonumber \\
 &=&
\frac{N\delta^{ab}g^2}{16\pi^2}\left[\frac{5}{3}\left(p^2 g_{\mu\nu}
- p_\mu p_\nu\right) - 3m^2 g_{\mu\nu}\right]\,.
\label{old.glghtotal}
\end{eqnarray}
We will now consider the loops based on cubic and quartic interactions 
between $B_{\mu\nu}$ and the gauge field $A_\mu$, for which the
vertices have been shown in Fig.~\ref{feynfig.ABB-AABB}.
\begin{figure}[htbp]
\centering
    \includegraphics[scale=0.085]{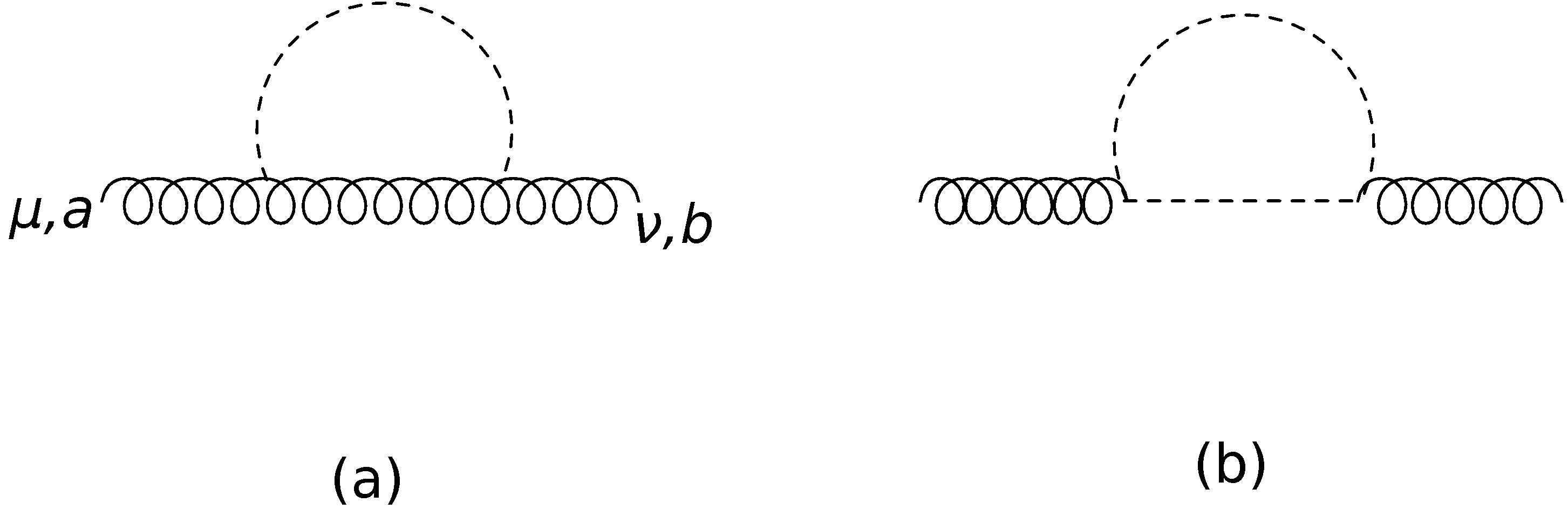} 
\caption{Loops formed by $AAB$ and $ABB$ couplings}
\label{oldfig.AAB-ABB-loop}
\end{figure}
The three point interactions $ABB$ and $AAB$ lead to the loop
diagrams shown in Fig.~\ref{oldfig.AAB-ABB-loop}. The coefficients
of $\frac{2}{\epsilon}$ from the loop integration corresponding to
them are respectively
 \begin{eqnarray}
 \Pi_{ab,\mu\nu, {\epsilon}}^{\ref{oldfig.AAB-ABB-loop}a} &=& 
\frac{N\delta^{ab}g^2}{16\pi^2}
3m^2 g_{\mu\nu}\,,\\ 
 \Pi_{ab,\mu\nu, {\epsilon}}^{\ref{oldfig.AAB-ABB-loop}b} 
&=& \frac{N\delta^{ab}g^2}{16\pi^2}\left[p^2 g_{\mu\nu} - p_\mu p_\nu 
+ 3m^2g_{\mu\nu}\right]\,.
 \end{eqnarray}
The propagator of the $B_{\mu\nu}$ field in the loop is the one
given in Eq.~(\ref{feyn.bprop}).  Adding the two contributions, we
get for the loops in Fig.~\ref{oldfig.AAB-ABB-loop}
\begin{eqnarray}
\Pi_{ab,\mu\nu,
  {\epsilon}}^{\ref{oldfig.AAB-ABB-loop}}=\Pi_{ab,\mu\nu,
  {\epsilon}}^{\ref{oldfig.AAB-ABB-loop}a} 
+\Pi_{ab,\mu\nu,
  {\epsilon}}^{\ref{oldfig.AAB-ABB-loop}b}=
\frac{N\delta^{ab}g^2}{16\pi^2} 
\left[p^2 g_{\mu\nu}-p_\mu p_\nu + 6 m^2 g_{\mu\nu}\right]\,.
\end{eqnarray}

\begin{figure}[thbp]
 \begin{center}
     \includegraphics[scale=0.08]{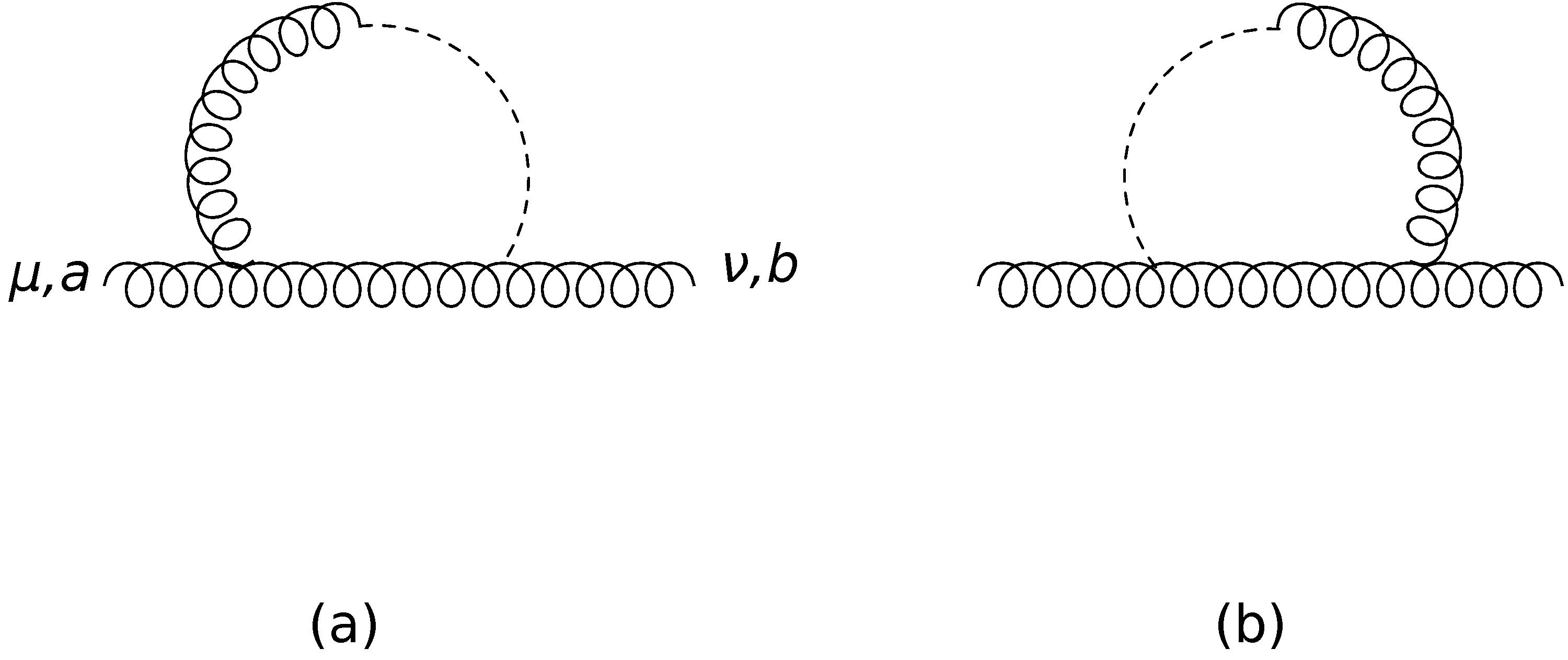} 
 \end{center}
\caption{Loop formed by $AAA$, $AB$ and $ABB$ couplings;}
  \label{oldfig.abloop2}
 \end{figure}

Next we consider the 1-loop diagrams generated by the cubic $AAA$
interaction at one vertex and the $AAB$ interaction at the other
vertex. These are shown in Fig.~\ref{oldfig.abloop2}, the
contributions from the two diagrams are equal.
The divergent part of the amplitudes corresponding to these
diagrams is
\begin{eqnarray}
\Pi_{ab,\mu\nu, {\epsilon}}^{\ref{oldfig.abloop2}a} &=&
-\frac{1}{2}\,\frac{N\delta^{ab}g^2}{16\pi^2}\,
\frac{9}{4}m^2g_{\mu\nu}
=\Pi_{ab,\mu\nu, {\epsilon}}^{\ref{oldfig.abloop2}b} 
\end{eqnarray}
The symmetry factor $\frac{1}{2}$ appears in
Fig.~\ref{oldfig.abloop2}(a) and Fig.~\ref{oldfig.abloop2}(b) since
these diagrams contain identical internal propagators of the gauge
field. In these diagrams, and in later ones, we take
Eq.~(\ref{feyn.bprop}) as the propagator of $B_{\mu\nu}$ whenever
an internal line of the loop contains the two point $AB$ vertex. As
before, we set $\eta=1\,.$ Hence the total contribution is
\begin{eqnarray}
\Pi_{ab,\mu\nu, {\epsilon}}^{\ref{oldfig.abloop2}} =
2\Pi_{ab,\mu\nu, {\epsilon}}^{\ref{oldfig.abloop2}a} 
= -\frac{N\delta^{ab}g^2}{16\pi^2}~~\frac{9}{4}m^2g_{\mu\nu}
\end{eqnarray}

\begin{figure}[hbtp]
 \begin{center}
    \includegraphics[scale=0.08]{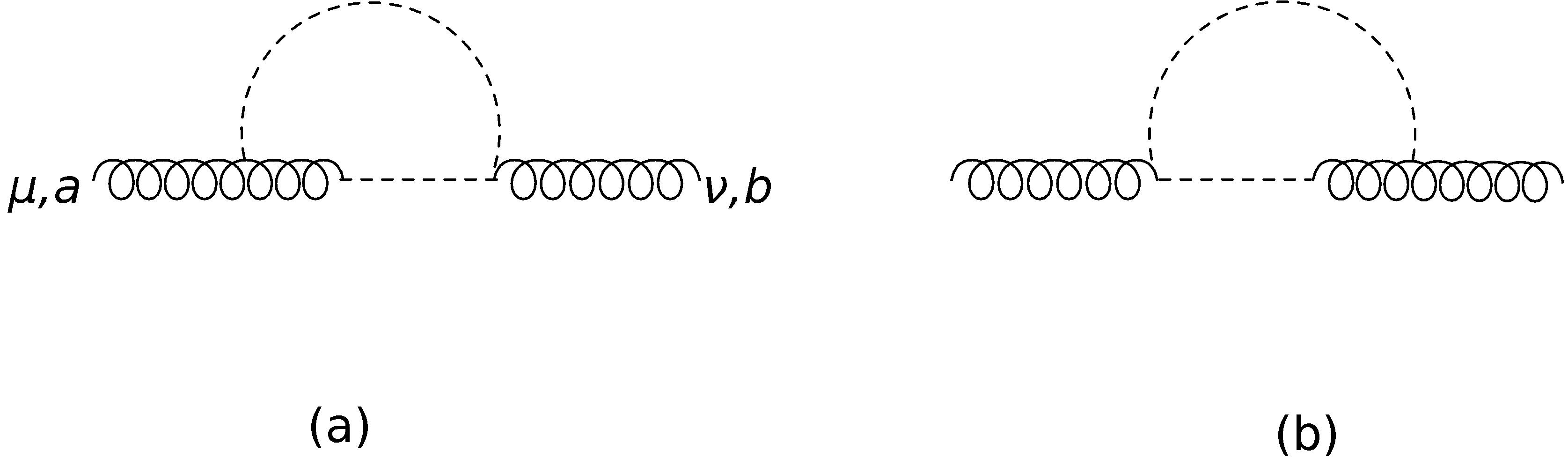} 
  \end{center}
\caption{Loops formed by $AAB$, $AB$ and $ABB$ couplings;}
  \label{oldfig.abloop3}
\end{figure}

Next we consider loops with the $AAB$ vertex at one end and the
$ABB$ coupling at the other end. We find the diagrams shown in
Fig.~\ref{oldfig.abloop3}.
The divergent parts corresponding to these diagrams are equal, and
are given by 
\begin{eqnarray}
\Pi_{ab,\mu\nu, {\epsilon}}^{\ref{oldfig.abloop3}a} &=&
-\frac{1}{2} 
\frac{N\delta^{ab}g^2}{16\pi^2} \frac{3}{2} m^2
g_{\mu\nu}=\Pi_{ab,\mu\nu, {\epsilon}}^{\ref{oldfig.abloop3}b} 
\end{eqnarray}
The factor $\frac{1}{2}$ for the diagrams in
Fig.~\ref{oldfig.abloop3}(a) and
Fig.~\ref{oldfig.abloop3}(b) is the symmetry factor for 
identical $B$-propagators in the loop.
 
\begin{figure}[htbp]
\begin{center}
    \includegraphics[scale=0.08]{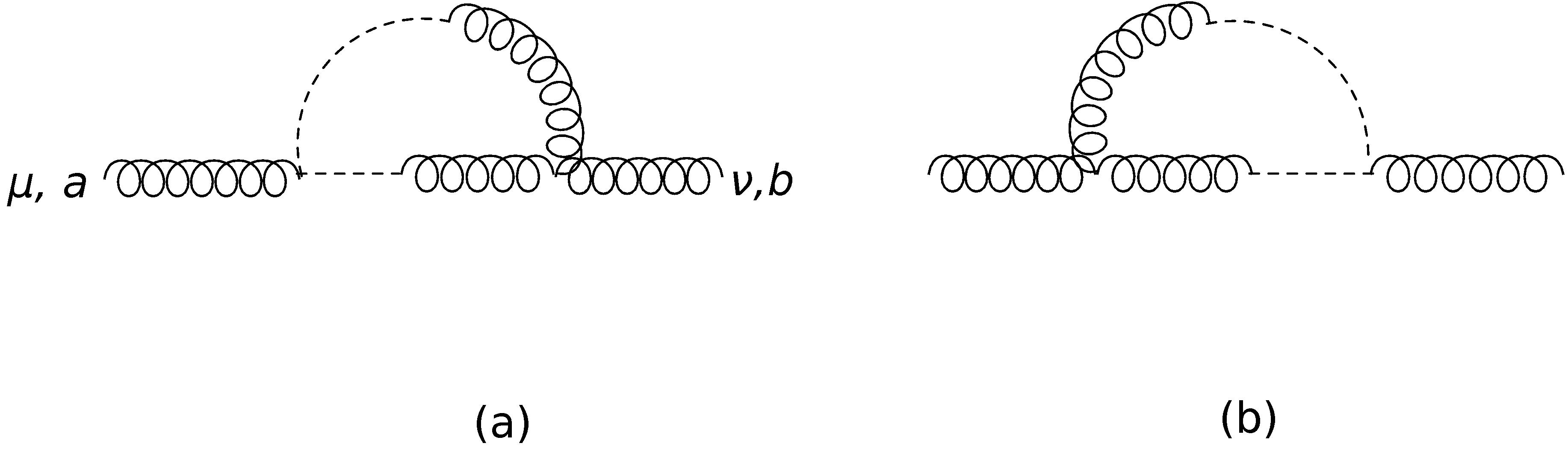} 
 \end{center}
\caption{Loop formed by $AAA$, $AB$, and $ABB$ couplings;}
  \label{oldfig.abloop4}
 \end{figure}

Using the couplings $AAA$, $AB$, and $ABB$ we get the diagrams
shown in Fig.~\ref{oldfig.abloop4} and both diagrams give the same
contribution,
%
\begin{eqnarray}
\Pi_{ab,\mu\nu, {\epsilon}}^{\ref{oldfig.abloop4}a}=
 -\frac{1}{4}\frac{N\delta^{ab}g^2}{16\pi^2}~~3m^2g_{\mu\nu}=
\Pi_{ab,\mu\nu, \epsilon}^{\ref{oldfig.abloop4}b} 
\end{eqnarray}
Now there are identical $A$-propagators as well as identical
$B$-propagators in each loop, so we pick up a symmetry factor of
$\frac12 \times\frac12 = \frac14\,.$

\begin{figure}[htbp]
\begin{center}
    \includegraphics[scale=0.08]{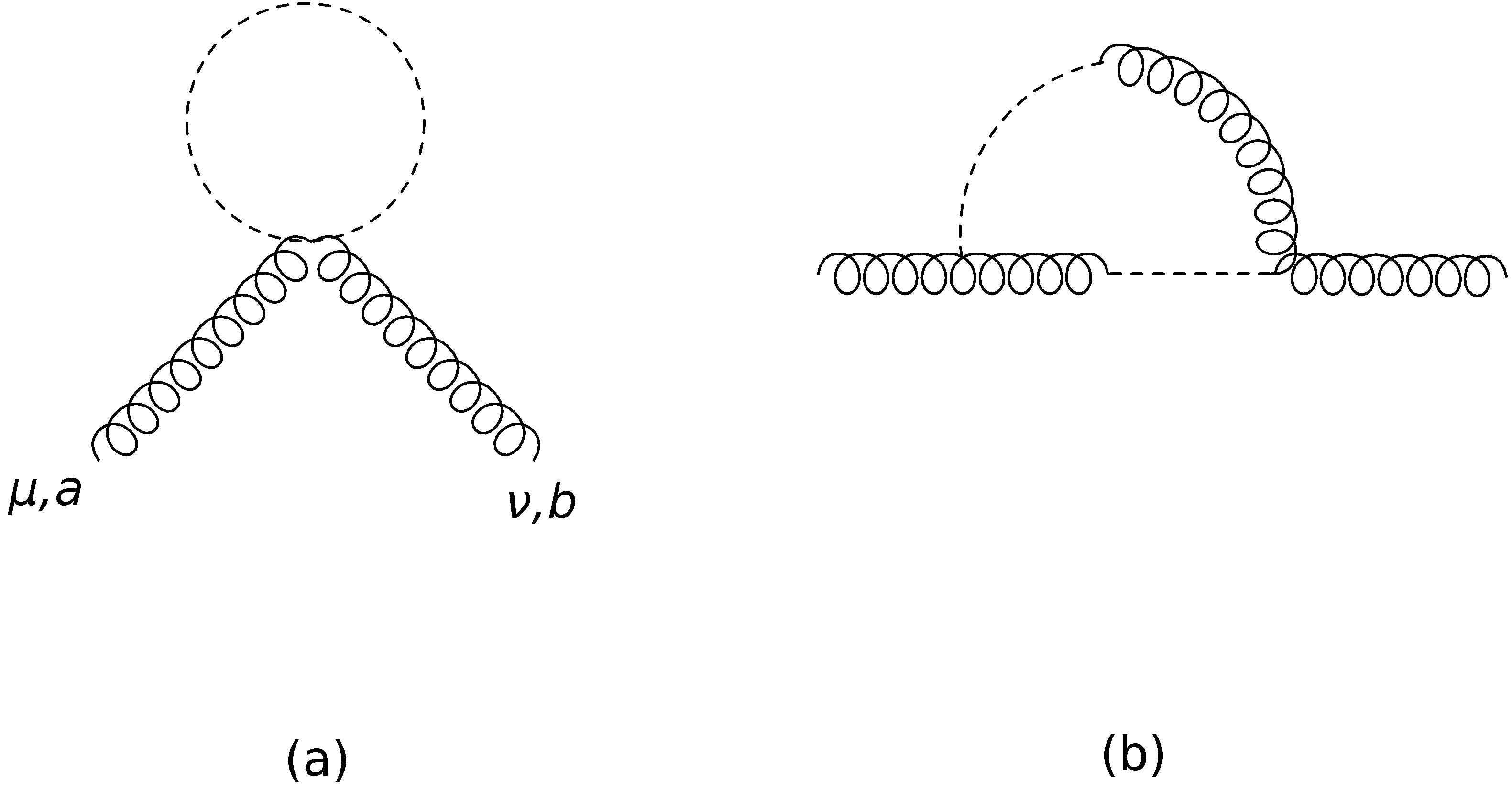} 
  \end{center}
\caption{(a) Loop formed by $AABB$ couplings; 
(b) Loop formed by $AAB$, $AB$, and $AAB$ couplings}
  \label{oldfig.abloop5}
\end{figure}

So the total contributions coming from the diagrams shown in 
Fig.~\ref{oldfig.abloop3} and Fig.~\ref{oldfig.abloop4} are
\begin{eqnarray}
\left(\Pi_{ab,\mu\nu, {\epsilon}}^{\ref{oldfig.abloop3}}
+\Pi_{ab,\mu\nu, {\epsilon}}^{\ref{oldfig.abloop4}}\right) 
&=&-\frac{N\delta^{ab}g^2}{16\pi^2}\,3m^2g_{\mu\nu}
\end{eqnarray}
Only two more one loop diagrams remain. The $AABB$ interaction in
the kinetic term of the $B_{\mu\nu}$ field leads to the diagram
shown in Fig.~\ref{oldfig.abloop5}(a), and the other diagram is
Fig.~\ref{oldfig.abloop5}(b), coming from the same interactions as
in Fig.~\ref{oldfig.abloop3}\,.
The diagram in Fig.~\ref{oldfig.abloop5}(b) is not divergent, so we 
can ignore it for our purposes. The diagram in
Fig.~\ref{oldfig.abloop5}(a) contributes  
\begin{eqnarray}
 \Pi_{ab,\mu\nu, {\epsilon}}^{\ref{oldfig.abloop5}} = 
-\frac{1}{2}\,\frac{N\delta^{ab}g^2}{16\pi^2}\,18 m^2 g_{\mu\nu}\,.
\end{eqnarray} 

Adding up the divergent parts corresponding to all the diagrams
from Fig.~\ref{oldfig.AAB-ABB-loop} -- Fig.~\ref{oldfig.abloop5}\,,
we find
\begin{eqnarray}
\Pi_{ab,\mu\nu, {\epsilon}}=\displaystyle \sum_{n=4}^9 
\Pi_{ab,\mu\nu, {\epsilon}}^{n}= \frac{N\delta^{ab}g^2}{16\pi^2}
\left[\frac{8}{3}\left(p^2 g_{\mu\nu} - p_\mu p_\nu \right)-
\frac{45}{4} m^2 g_{\mu\nu}\right]\,.
\label{old.fullpi}
\end{eqnarray}
%
\section{Beta function}\label{beta}
\begin{figure}[hbtp]
\begin{center}
    \includegraphics[scale=0.068]{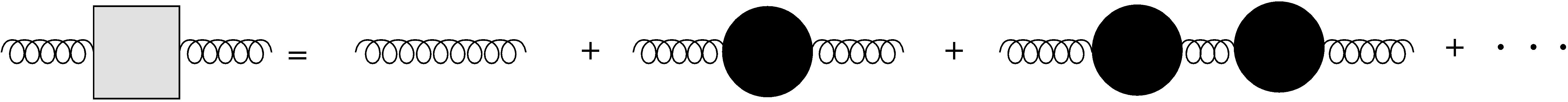} 
  \end{center}
\caption{Exact gluon propagator at one loop}
  \label{beta.loopsum2}
\end{figure}

The exact propagator of the gauge field at the one-loop level is
calculated by summing over insertions of the one-loop diagrams as
in Fig.~\ref{beta.loopsum2}\,,
corresponding to the equation 
\begin{eqnarray}
\tilde{\Delta}_{\mu\nu}=iD_{\mu\nu} +iD_{\mu\alpha}
i\Pi^{\alpha\beta}iD_{\beta\nu} +iD_{\mu\alpha}i\Pi^{\alpha\beta}
iD_{\beta\gamma}i\Pi^{\gamma\delta}iD_{\delta\nu}+\cdots\,,
\label{beta.tDelta}
\end{eqnarray}
where we have suppressed the gauge indices. The solid blob is what  
we have calculated so far in Sec.~\ref{old}, the sum of one-loop
diagrams which contribute to the $A$-propagator. 

Looking at the one loop corrections we have calculated, we see that 
the general structure of the correction $\Pi^{\alpha\beta}$ is
\begin{eqnarray}
\Pi^{\alpha\beta}(k) = \pi_1(k^2, m^2)(g^{\alpha\beta}k^2-k^\alpha
k^\beta) +\pi_2(k^2,m^2) m^2 g^{\alpha\beta}\,.
\label{beta.Piab}
\end{eqnarray}
This form is misleading, however. The pole in the propagator of
Eq.~(\ref{feyn.aprop}) came from an infinite sum over massless
propagators, with the two-point interaction inserted in
between. If we want to find the correction in the pole, we ought
to consider loop corrections to the two-point vertex of
Fig.~\ref{feynfig.AB-AAB}(a), as well as to the $B$-propagator, in
addition to the $A$ propagator. But it is not necessary to do that
for our purpose. Our original goal was to calculate the
$\beta$-function of the gauge coupling constant, let us see how 
that is related to the calculations we have done so far.

We started our calculations from the renormalized Lagrangian
density in Eq.~(\ref{feyn.lag}). 
The Lagrangian density for the counterterms is thus
\begin{eqnarray}
\nonumber \mathscr{L}_{ct}&=& 
(Z_3-1) \frac{1}{4}F^{\mu\nu}_a F_{\mu\nu}^a
+(Z_1-1) \frac{1}{12}H^{\mu\nu\lambda}_a H_{\mu\nu\lambda}^a
+(Z_m-1) \frac{m}{4}\epsilon ^{\mu\nu\rho\lambda}F_{\mu\nu}^a
B_{\rho\lambda}^a  \\ \nonumber  && + (Z'_{2}-1)\partial_\mu
\bar{\omega}_a\partial^\mu \omega^a  
- (Z'_1-1)g f_{bca} A_\mu^b\partial^\mu \bar{\omega}^a\omega_c
-(Z_3-1)\frac{1}{2\xi}(\partial_\mu A^\mu_a)^2\\ 
&& -(Z_1-1)\frac{1}{2\eta}(D_\mu B^{\mu\nu})^2\,.
\label{beta.lct}
\end{eqnarray}
The counterterms crucial to our calculations are the ones related
to gauge invariance under SU(N) gauge symmetry. It is easy to check
that the counterterm Lagrangian of Eq.~(\ref{beta.lct}) is
invariant under a renormalized BRST symmetry, which implies that
the SU(N) gauge symmetry remains unbroken. In particular, in the
chosen gauge, the Slavnov-Taylor identity coming from the BRST
transformation is the same as for usual (massless) Yang-Mills
theory, showing that only the transverse part of the propagator
gets a correction\footnote{We thank the anonymous referee for
  asking for a clarification on this point.}. This supports our
earlier comment that loop corrections to the two-point coupling as
well as to the $B$-propagator have to be taken into account in
order to calculate $\pi_2$\,.

On the other hand, the $\beta$-function for the gauge coupling
constant is affected by the $B$-field only through the calculation
of the renormalization constant $Z_3\,,$ which is related to
$\pi_1\,.$ From Eqs.~(\ref{old.fullpi}) and (\ref{beta.lct}), we
get
\begin{eqnarray}
Z_3=1+\frac{N g^2}{16\pi^2}\frac{8}{3}\left[\frac{2}{\epsilon} - 
   \ln\frac{m^2}{\mu^{2}}\right] \,,
\label{beta.Z3}
\end{eqnarray}
where $\mu$ is the subtraction point, and we have ignored a 
constant, independent of $\mu$ and $m$, in the second term. 

In order to calculate the $\beta$-function for the gauge coupling
constant, we need the relation between the gauge coupling constant
at momentum scale $\mu$ with the bare coupling constant. If we add
$\mathscr{L}_{ct}$ with the renormalized Lagrangian given in
Eq.~(\ref{feyn.lag}), we find the bare Lagrangian density,
\begin{eqnarray}
\nonumber \mathscr{L}_{\mathscr{B}}&=&
\mathscr{L}+\mathscr{L}_{ct}\\ \nonumber &=& -\frac{1}{4}F^{\mu\nu
  a}_\mathscr{B} F_{\mathscr{B}{\mu\nu a}}+
\frac{1}{12}H^{\mu\nu\lambda a}_\mathscr{B}
H_{\mathscr{B}{\mu\nu\lambda 
  a}}+ \frac{m_\mathscr{B}}{4}\epsilon
^{\mu\nu\rho\lambda}F_{\mathscr{B} {\mu\nu a}}
B_{\mathscr{B}{\rho\lambda  a}} \\ && + \partial_\mu
\bar{\omega}_{\mathscr{B}a} \partial^\mu \omega^a_\mathscr{B} -
g_\mathscr{B} f_{bca} A_{\mathscr{B}\mu}^b\partial^\mu
\bar{\omega}^a_\mathscr{B}~\omega_{\mathscr{B}c}\,,
\end{eqnarray}
where 
\begin{eqnarray}
A_\mathscr{B} &=& Z^{\frac{1}{2}}_3 A\\
B_\mathscr{B} &=& Z^{\frac{1}{2}}_1 B\\
\omega_\mathscr{B} &=& {Z'_{2}}^{\frac{1}{2}}\omega\\
m_\mathscr{B} &=& \frac{Z_m}{Z^{\frac{1}{2}}_3Z^{\frac{1}{2}}_1}m\\
g_{\mathscr{B}} &=&\frac{Z'_1}{Z^{\frac{1}{2}}_3 Z'_{2}}g\,,
\label{beta.g_bare}
\end{eqnarray}
and the subscript $\mathscr{B}$ denotes bare fields and coupling
constants.  

\begin{figure}[tbp]
\begin{center}
 \includegraphics[scale=0.07]{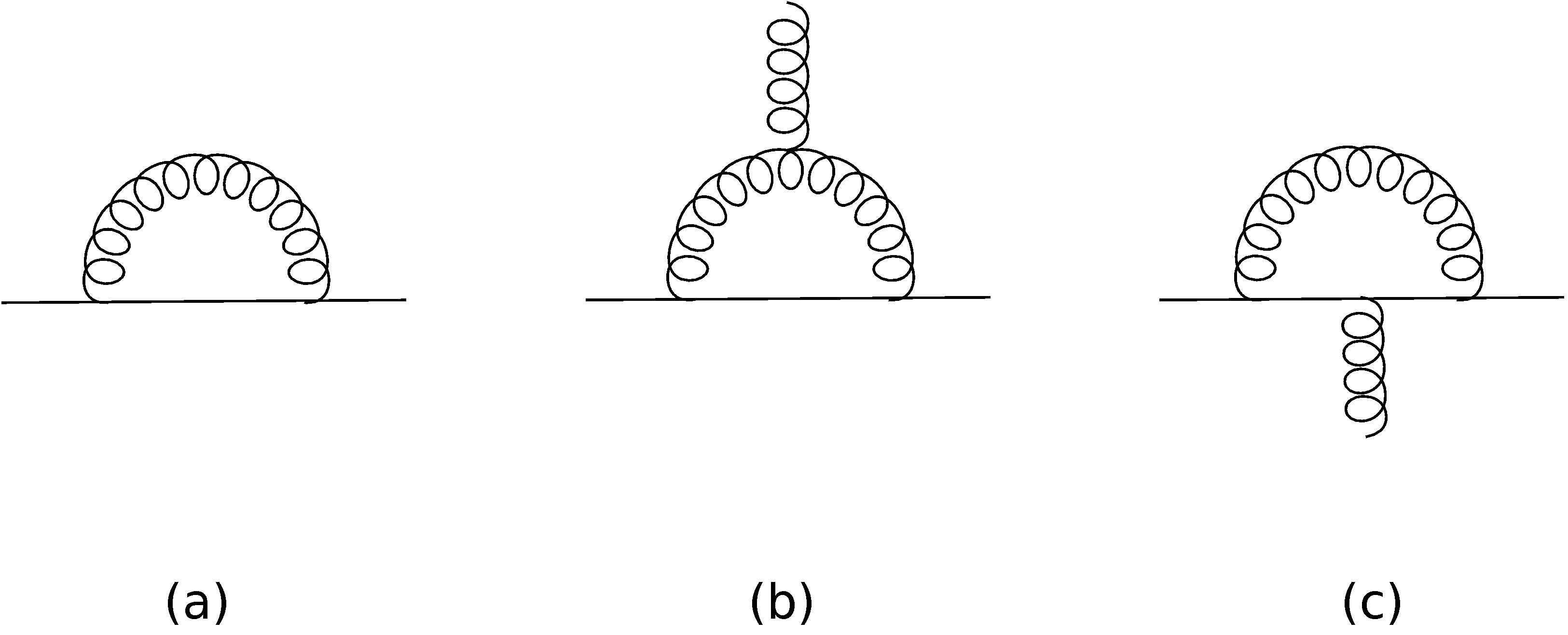} 
\caption{One loop contributions to $Z'_1$ and $Z'_2$}
  \label{fig:ghcorr}
  \end{center}
\end{figure}

We consider the diagram Fig.~\ref{fig:ghcorr}(a) for the the one
loop correction for the ghost propagator. It can be checked
explicitly that the divergent part of this diagram, and thus
$Z'_2$\,, is unaffected by a massive pole in the gauge field
propagator.  Similarly it can be checked by explicit calculation
that the mass does not affect the one loop corrections to the
gauge-ghost vertex, and thus $Z'_1$\,. The relevant diagrams are
shown in Fig.~\ref{fig:ghcorr}(b) and \ref{fig:ghcorr}(c).

The reasons are easy to understand by looking at the diagrams.
The $A^\mu\bar\omega\omega$ interaction has the tree-level vertex
rule 
\begin{eqnarray}
iV^\mu_{abc}=- g f_{bca}p^\mu\,,
\end{eqnarray}
where $p^\mu$ is the incoming momentum carried by the ghost field.
The external momentum does not enter the loop integration, so the
one loop divergence remains logarithmic as in usual Yang Mills
gauge theory, and the pole does not contribute to the divergent
part of $Z'_2$\,. Similarly, the vertex correction contains three
internal lines with three momentum dependent vertices, but the
divergent part of the loop integral comes from the leading power of
the internal momentum.  As a consequence, the loop divergence is
also logarithmic.  So the divergent parts of the renormalization
factors are the same as in pure Yang Mills theory,
\begin{eqnarray}
Z'_1 &=& 1-\frac{Ng^2}{32\pi^2}\left(\frac{2}{\epsilon} - 
\ln\frac{m^2}{\mu^2}\right) \label{beta.Z'1} \\
Z'_2 &=& 1+\frac{Ng^2}{32\pi^2}\left(\frac{2}{\epsilon} -
\ln\frac{m^2}{\mu^2}\right)\,.\label{beta.Z'2}
\end{eqnarray}

We can now calculate the beta function of $\alpha =
\dfrac{g^2}{4\pi}$  
using Eq.s~(\ref{beta.g_bare}), 
(\ref{beta.Z3}), (\ref{beta.Z'1}) and (\ref{beta.Z'2}), 
\begin{eqnarray}
\beta(\alpha)= \frac{\partial g(\mu)}{\partial\log\mu}
=-\frac{14}{3} N\frac{\alpha^2}{2\pi} \,.
\end{eqnarray}
So the theory remains asymptotically free in spite of the
appearance of a massive pole in the gauge field propagator due to 
the topological coupling.

\section{Comments and conclusion}
\begin{figure}[h]
\begin{center}
    \includegraphics[scale=0.25]{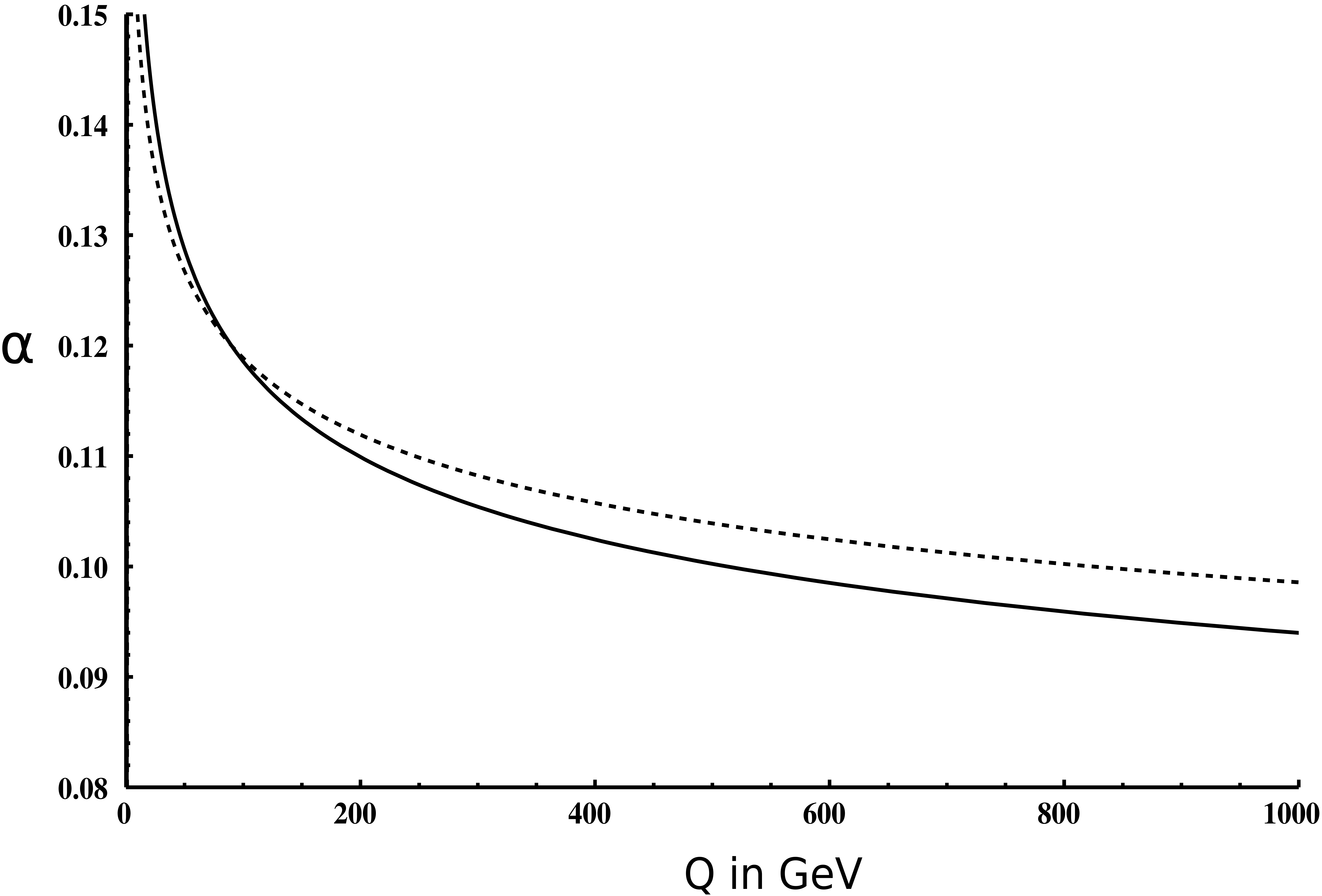} 
  \end{center}
\caption{The flow of $\alpha_s$ for ordinary (dotted) and
  topologically massive (solid) SU(3) gauge theory.} 
  \label{fig:runningcc}
\end{figure}

The main result of this paper is that non-Abelian gauge theory
coupled to an antisymmetric tensor field is both short range and
asymptotically free. We can think of the antisymmetric tensor as a
minimally coupled matter field for the calculations done in this
paper.  In general, interaction with scalar and fermion fields has
a screening effect for color charge, and the gauge coupling
constant becomes flatter with respect to pure non-Abelian gauge
theory, as the fields contribute positively to the beta function.
But the interactions of the gauge field with the tensor field
causes the coupling constant to fall away more steeply, as the
additional contribution to the beta function is negative.  We show
in Fig.~\ref{fig:runningcc} the running of $\alpha$ as a function
of $Q$ using the relation
\begin{eqnarray}
\alpha(Q^2)=\frac{\alpha(\mu^2)}{1+N\frac{c}{\pi}\alpha(\mu^2)
\ln\frac{Q^2}{\mu^2}}\,,
\label{end.alpha}
\end{eqnarray} 
where $N=3\,,$ and $c=\frac{11}{12}$ for the pure massless
Yang-Mills gauge field and $c=\frac{14}{12}$ for the topologically 
massive field considered in this paper.
For the plot, we have taken  $\mu^2=M_Z^2\,,$ with
$\alpha(M_Z^2)=0.12$~\cite{Beringer:1900zz}.

We did not consider quantum corrections to the propagator of the
tensor field or to the two-point vector-tensor coupling.
Consequently, we cannot make any comment on the renormalization of
$m$ in this theory. Dynamically generated gluon mass vanishes at
short distances~\cite{Cornwall:1981zr}.  For the model we have
considered here, the propagator has a pole and is well-behaved at
short distances, in fact the gauge boson propagator in
Eq.~(\ref{feyn.aprop}) behaves exactly like one with a dynamically
generated mass~\cite{Aguilar:2009ke}. We are unable to say at this
point how $m$ behaves at short distances in this model, but the
form of the propagator, as well as the asymptotic behavior of
$\alpha_s\,,$ indicates that the theory may be renormalizable, as
has been argued algebraically elsewhere~\cite{Lahiri:1999uc}.


\end{document}